\documentstyle[preprint,tighten,eqsecnum,aps,floats,psfig,epsfig]{revtex}

\begin{document}
\draft
\title{
Comment on ``Spurious fixed points in frustrated magnets," 
cond-mat/0609285
}
\author{
Andrea Pelissetto,$^1$ and Ettore Vicari$\,^2$ }
\address{$^1$ Dip. Fisica dell'Universit\`a di Roma ``La Sapienza" \\
and INFN, P.le Moro 2, I-00185 Roma, Italy}
\address{$^1$
Dip. Fisica dell'Universit\`a di Pisa
and INFN, V. Buonarroti 2, I-56127 Pisa, Italy}
\address{
\bf e-mail: \rm 
{\tt Andrea.Pelissetto@roma1.infn.it},
{\tt vicari@df.unipi.it}
}

\date{\today}

\maketitle

\begin{abstract}
  We critically discuss the arguments reported in cond-mat/0609285 by B.
  Delamotte, Yu. Holovatch, D. Ivaneyko, D. Mouhanna, and M. Tissier.  We show
  that their conclusions are not theoretically founded.  They are contradicted
  by theoretical arguments and numerical results. On the contrary,
  perturbative field theory provides a robust evidence for the existence of
  chiral fixed points in $O(2)\otimes O(N)$ systems with $N\ge 2$. The
  three-dimensional perturbative results are consistent with theory and with
  all available experimental and Monte Carlo results. They provide a
  consistent scenario for the critical behavior of chiral systems.
\end{abstract}




In Ref.~\cite{DHIMT-06} the authors provide what they think is 
evidence against the presence of a stable fixed point (FP) in 
$O(2)\otimes O(N)$ models for $N=2,3$, contradicting
the results of Refs.~\cite{PRV-01,CPPV-04} obtained by the analysis
of high-order perturbative series (see, e.g., Ref.~\cite{ZJ} for 
a discussion of the method).
In this Comment we show that 
their arguments are either theoretically incorrect or contradictory. 

\vskip 0.3truecm

{\em Fixed points in four dimensions.}
The main argument against the presence of a stable FP for $N=2,3$ is 
based on the claim that a {\em physical} FP of a given Hamiltonian must 
survive up to $d=4$ (i.e. up to $\epsilon \equiv 4-d = 0$) and become 
the Gaussian FP in this limit. 
There is no theoretical justification 
for this requirement: the authors of Ref.~\cite{DHIMT-06} do not
provide any motivation, they simply state it must be so. Note that this 
condition is very restrictive and contradicts several well-accepted 
theoretical results. For $\epsilon \ll 1$, the chiral
FP $C_+$ exists only for $N\gtrsim 21.8$. Thus, according to this 
criterion, all FPs with $N \lesssim 
21.8$ should be considered as {\em spurious}! This is in contradiction
with the results obtained in all approaches: indeed, for $N > 6$ 
all methods agree and predict a stable FP. 
We must observe that the perturbative results of 
Ref.~\cite{CPPV-04} find no difference between 
the FPs with $N\gtrsim 6$ (on whose existence there is no discussion)
and those with $N\lesssim 6$: the FP position varies smoothly
with $N$ and $d$. Thus, the difference between the perturbative 
results and those obtained 
by using the functional renormalization group (FRG) \cite{TDM-00,TDM-03,DMT-04} 
is only quantitative: the two methods disagree on the behavior of the 
function $N_c(\epsilon)$ or of its inverse\footnote{
The function $N_c(\epsilon)$ and its inverse $\epsilon_c(N)$ are
defined, e.g., in Ref.~\cite{CPPV-04}, Sec.~II.D. For each $N$,
the function  $\epsilon_c(N)$
is defined so that a stable FP exists for $\epsilon > \epsilon_c(N)$,
while no FP exists in the opposite case $\epsilon < \epsilon_c(N)$.
The $\epsilon$ expansion predicts $\epsilon_c(N) = 0$ for $N\gtrsim 21.8$.
Perturbation theory (Refs.~\cite{PRV-01,CPPV-04}) gives
$\epsilon_c(N) < 1$ for all $N\ge 2$.}
$\epsilon_c(N)$ 
for $\epsilon \approx 1$ (see the discussion in Sec.~II.D of 
Ref.~\cite{CPPV-04}).
But there are no conceptual differences as the authors of Ref.~\cite{DHIMT-06}
apparently imply. 

The behavior observed here 
is analogous to that found in the Ginzburg-Landau model of superconductors.
Close to 4 dimensions a FP is found only for $N\gtrsim 365.9$ ($N$ is the 
number of components of the scalar field) \cite{HLM-74}. The criterion proposed 
in Ref.~\cite{DHIMT-06} would thus predict a first-order transition 
for the physical case $N=1$, contradicting experiments \cite{GN-94},
and also general duality arguments \cite{kleinertbook},
FRG calculations \cite{Betal-96}, and 
Monte Carlo simulations \cite{MHS-03}.
Fixed-dimension calculations are instead consistent with the 
existence of a stable FP \cite{HT-96,FH-96}.

As we already said, we do not think that the behavior in four dimensions
is of any relevance for the three-dimensional theory. There is instead
another condition that is crucial: the three-dimensional FP must be 
connected by the three-dimensional renormalization-group flow 
to the Gaussian FP \cite{ZJ,ID}. If this is the case,
at least for the massive zero-momentum (MZM) scheme, one can give a 
rigorous nonperturbative definition of the renormalization-group flow and
of all quantities that are computed in perturbation theory.  
One may consider weakly coupled models or models with medium-range 
interactions, as discussed in Sec.~III.B of Ref.~\cite{CPPV-04}. 
These models are defined on the lattice and as such are well defined 
nonperturbatively. In a well-defined limit (see 
Sec. III.B of Ref.~\cite{CPPV-04}) long-range quantities 
(correlation length, susceptibilities, etc.) have the {\em same}
perturbative expansion as the corresponding quantities in the 
continuum theory in the MZM scheme. 
The existence of a stable FP is equivalent to the 
existence of a second-order transition for small but finite values of 
the bare quartic coupling constants (weakly coupled models) or for large but
finite values of the interaction range (medium-range systems).
Note that everything is defined nonperturbatively and rigorously
in three dimensions and there is 
no need of invoking the existence of a four-dimensional FP. 

\vskip 0.3truecm

{\em The chiral fixed points for $d\to 4$.}
In Ref.~\cite{DHIMT-06} it was claimed that for $N=2,3$ one can follow 
the FP $C_+$ up to $d=4$. This statement is incorrect and is based on an 
incorrect use of the conformal-mapping method. To explain this point 
let us review the results of the semiclassical analysis for the 
Borel singularities. For $u\ge 0$ and $v \le 2 u$ the perturbative series
are Borel summable. Thus, under the assumption that all singularities
lie on the negative real axis in the Borel plane, the conformal-mapping 
method transforms the original asymptotic series into a convergent one. 
If  $2 u \le v \le 4 u$ the perturbative series are not Borel summable:
there is a singularity on the positive real axis. However, in this range 
of values of $v$ the Borel singularity that is closest to the origin is still 
on the negative real axis. Thus, the conformal-mapping method still takes 
into account the large-order behavior. Though the resulting resummed 
series is only asymptotic, its coefficients
should increase slower than those of the original one. 
For these values of $v$ one may expect to obtain reasonable results 
out of six- or five-loop perturbation theory, 
though not so precise as those that are obtained 
in $O(N)$ models. If $v\ge 4 u$ the leading singularity is on the 
positive real axis and thus the conformal-mapping method cannot be used:
the function $\omega(u;z)$ (defined in Eq.~(2) of Ref.~\cite{DHIMT-06})
becomes imaginary for $u = -1/a$ if $a$ is 
negative. For this reason, as discussed in Ref.~\cite{CPPV-04}, one cannot 
follow the FPs that exist for $N=2$ and $3$ for $d \gtrsim 3.3$. 
Probably, Fig.~4 of Ref.~\cite{DHIMT-06} has been obtained by always 
using $a = u/2$, which means that for $v \ge 4 u$ the resummation does 
not take into account the leading large-order behavior: the 
{\em resummed} perturbative series diverge {\em exactly} as the original ones.
Unresummed perturbative series are not very predictive and the most 
accurate results (we use the word accurate in the sense of least difference 
between estimate and exact result, as usual in the context of asymptotic 
series\footnote{
We remind the reader that, given a quantity ${\cal O}(x)$ that has
an asymptotic expansion ${\cal O}(x)\approx \sum_n a_n x^n$, for each
$x$ there is an optimal $k(x)$ such that the least error
${\rm min}_h|{\cal O}(x) - \sum_{n\le h} a_n x^n|$ is obtained by taking
$h = k(x)$. For a convergent series $k(x)=\infty$; if the series is
divergent $k(x)$ is finite. In the case of perturbative expansions
of the $O(N)$ models, $k(g) = 2$ for $g$ close to the FP:
the best results are obtained at two loops!
})
are obtained at very low orders. For instance,
in the absence of a Borel resummation,
the standard $\epsilon$ expansion for $O(N)$ models gives the most accurate
results at two loops \cite{BGZN-73}: inclusion of higher-order terms
worsens the estimates and indeed unresummed five-loop estimates
have no relation with the correct asymptotic result. 
The same occurs in the MZM scheme. 
Thus, the results of Ref.~\cite{DHIMT-06} for the 
existence of FPs for $N=2,3$ in the intermediate region 
$3.3\le d \le 4$ cannot be trusted. 

\vskip 0.3truecm
{\em Numerical results.}
The authors of Ref.~\cite{DHIMT-06} claim that the existence of a stable
FP contradicts presently available numerical results. Instead,
there are no contradictions: All 
numerical and experimental results are consistent with the predictions
of perturbative field theory. First, let us remind the reader that the 
existence of a stable FP does not imply that {\em all} systems
with the given symmetry undergo a second-order phase transition. 
The transition is continuous only if the system is in the basin of attraction 
of the stable FP.
For instance, in systems with a scalar order parameter one may observe
a first-order transition or a second-order transition belonging to the 
Ising universality class; with an appropriate tuning of the parameters
it is also possible to have a tricritical point. 
First-order transitions in scalar systems are perfectly compatible with 
the existence of the Ising universality class!
Thus, the quoted results \cite{Itakura-03,PSDMT-04,BSPM-06}
are {\em irrelevant} for the discussion: the first-order nature of the 
transition in XY stacked antiferromagnets (not to be confused 
with the easy-plane antiferromagnets studied experimentally,
see the discussion in Sec. V of Ref.~\cite{CPPV-04}) is consistent both with
the presence and with the absence of an $N=2$ chiral FP. 
On the other hand, the results of Ref.~\cite{CPPV-04} and of 
Ref.~\cite{PS-03}---they find continuous transitions 
for $N=2$ and $N=3$, respectively---are only consistent with the presence
of a stable FP and thus do not support the scenario of Ref.~\cite{DHIMT-06}. 
Thus, at variance with the claim of Ref.~\cite{DHIMT-06},
numerical results are {\em not} consistent with the scenario
that all chiral systems with $N=2$ and $N=3$ undergo first-order transitions.

\vskip 0.3truecm
{\em Experimental results.}
Ref.~\cite{DHIMT-06} states that experiments contradict 
the perturbative results. Again, we must stress that this is not the case. 
First, we must note that the results of Ref.~\cite{QHPP-06} 
cited in Ref.~\cite{DHIMT-06} are perfectly
consistent with perturbation theory. We discussed in detail 
easy-axis systems in Ref.~\cite{CPV-05} and showed two possible 
phase diagrams compatible with perturbation theory, see Fig. 3 of 
Ref.~\cite{CPV-05}. The results of Ref.~\cite{QHPP-06} correspond 
exactly to phase diagram (b) reported in Fig.~3 of Ref.~\cite{CPV-05}.  
As for easy-plane 
systems, all experiments observe continuous transitions, and thus they 
are compatible with the perturbative results (but not with those of 
Refs.~\cite{TDM-00,TDM-03,DMT-04}). In any case, note that perturbation theory
does not predict easy-plane systems (or any model) 
to have a continuous transition. 
It only predicts that, if the transition is continuous, it should 
belong to the chiral universality class. In Refs.~\cite{TDM-00,TDM-03,DMT-04}
it was remarked that, even if experiments observe continuous transitions,
the measured critical exponents in some cases 
do not satisfy rigorous inequalities, for instance $\beta \ge \nu/2$.
In Ref.~\cite{CPPV-04} we showed that this may be explained 
by corrections to scaling. Due to the ``focus"-like nature of the 
FP \cite{CPS-02-03} the approach to criticality may be quite complex.
Effective exponents may even change nonmonotonically, at variance with what
occurs in $N$-vector models (for the effective exponents in 
$O(N)$ models, see Refs.~\cite{BB-02,PRV-99}). 

In conclusion, the statement of Ref.~\cite{DHIMT-06} on the experiments is
biased and not justified by the experimental results.

\vskip 0.3truecm
{\em The cubic model.}
The authors of Ref.~\cite{DHIMT-06} also discuss the cubic model defined in 
their Eq.~(4). They assume the {\em well-established} scenario provided by 
the $\epsilon$ expansion:
no FP is present in the domain $v<0$. Then, they analyze the perturbative 
series for the model in the 
minimal-subtraction scheme without $\epsilon$-expansion
($3d$-$\overline{\rm MS}$ scheme) and find some evidence for a FP with 
$v < 0$. Since this FP is not predicted by the $\epsilon$ expansion
they conclude that this FP is spurious. Hence---they conclude---perturbation 
theory cannot be trusted.

Again, the presentation is strongly biased. 
First, it is not clear to us which results support the claim 
that the physics of systems with cubic anisotropy is {\em well established}.
Indeed, all references that are cited and that discuss the behavior 
for $v < 0$ {\em assume} that the $\epsilon$ expansion is predictive 
in three dimensions. 
But, of course, one cannot take this for granted in this discussion, 
since the debate is exactly on the quantitative predictivity of 
the $\epsilon$ expansion.
Beside the $\epsilon$-expansion investigations, there are only 
a few numerical works for the antiferromagnetic four-state Potts 
model on a cubic lattice \cite{USO-89,Itakura-99,YO-01}, 
whose critical behavior should 
be described by the $N=3$ cubic model with $v < 0$ \cite{BGJ-80-82}.
Contrary to the claim of Ref.~\cite{DHIMT-06}, all numerical 
results are consistent with a continuous transition: 
at present there is no evidence of first-order transitions.
Thus, the behavior of this class of systems is not at all 
{\em well established}, but appears as rather controversial. 
Therefore, it is worthwhile to check whether 
a FP with $v < 0$ exists or not, without any {\em a priori}
preconception.

\begin{figure}[tb]
\centerline{\psfig{width=10truecm,angle=-90,file=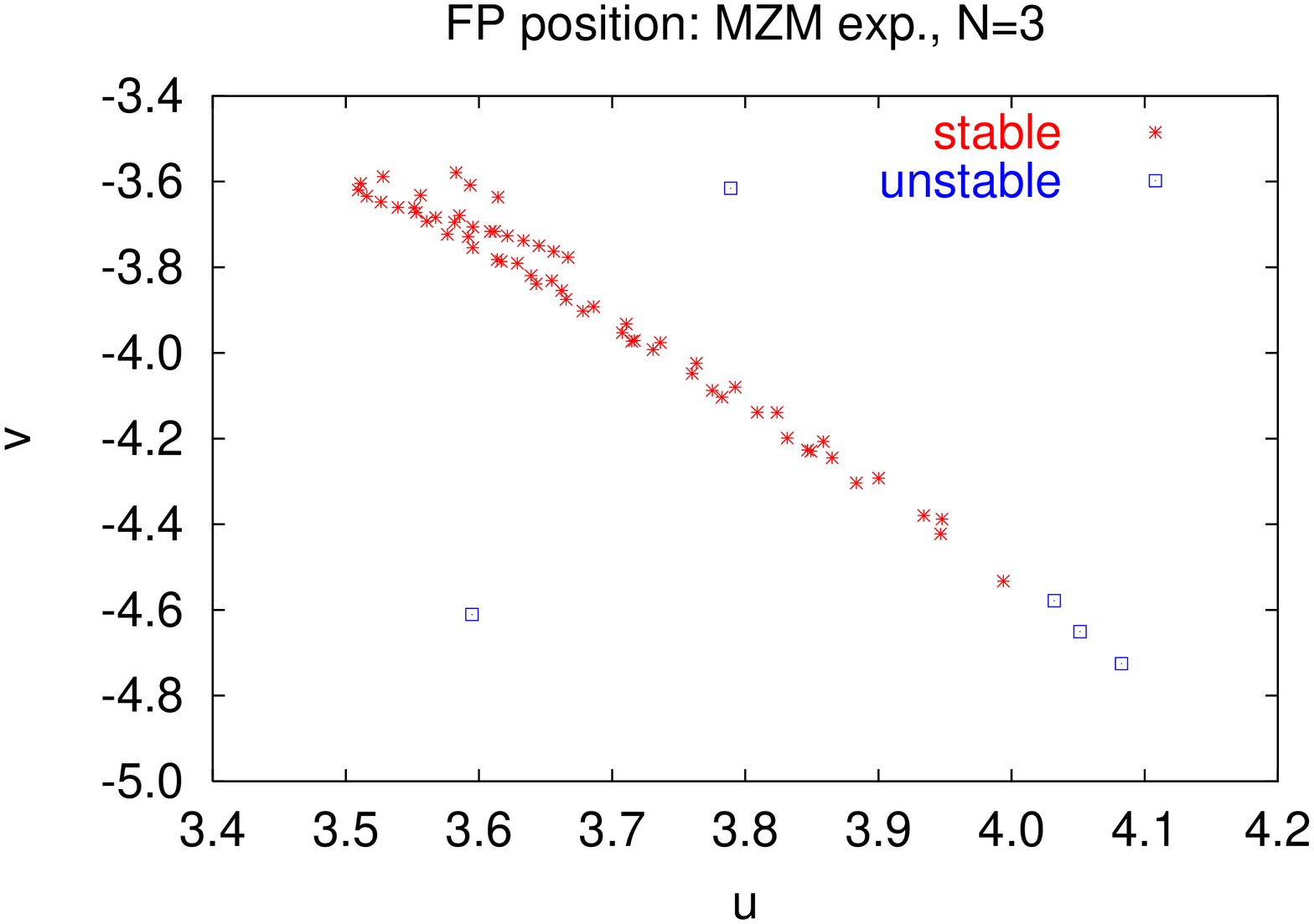}}
\centerline{\psfig{width=10truecm,angle=-90,file=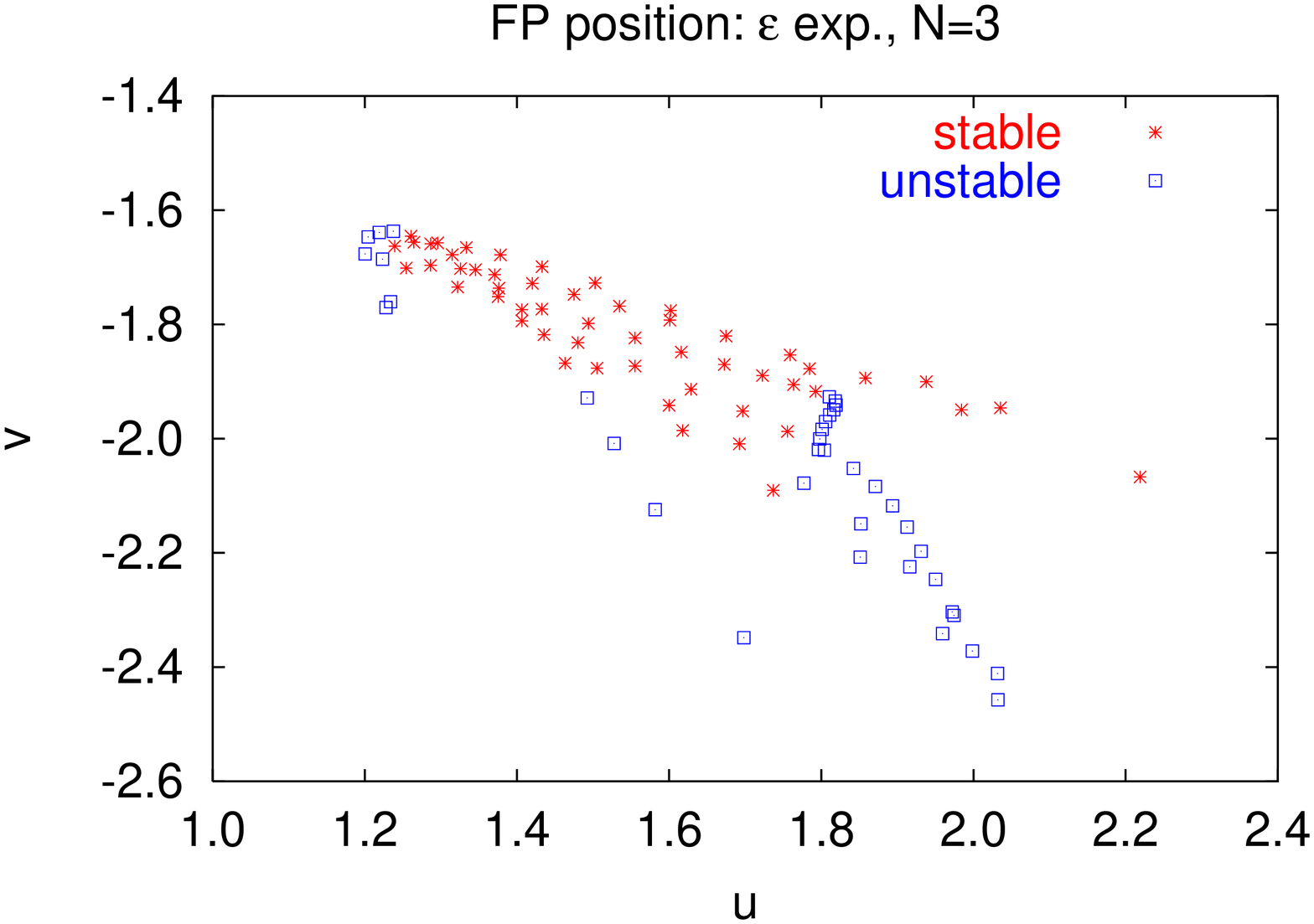}}
\vspace{2mm}
\caption{Perturbative results for the cubic model with $N=3$
and $v < 0$: Fixed-point position. Above we report the 
MZM results, below the $3d$-${\overline{\rm MS} }$ results.}
\label{cubico}
\end{figure}

For this purpose we have repeated the analysis of the 5-loop 
perturbative series for the 
cubic model in the $3d$-$\overline{\rm MS}$ scheme
discussed in Ref.~\cite{DHIMT-06} and we have also analyzed 
the 6-loop perturbative series of Ref.~\cite{CPV-00} in the MZM
scheme. We focus on $v < 0$ and $N=3$. 
For each $b$ and $\alpha$ (defined in Ref.~\cite{DHIMT-06} or, equivalently
in Ref.~\cite{CPV-00}) we resum the $\beta$ functions,\footnote{
In the $3d$-$\overline{\rm MS}$ scheme we resum
$f_z(u,v) = (\beta_z(u,v) + z)/z^2$, $z=u,v$ as in Ref.~\cite{DHIMT-06}.
Similar results are obtained by resumming $\beta_z(u,v)/z$.
In the MZM scheme we resum $\beta_z(u,v)/z$, as in Ref.~\cite{CPV-00}.} 
look for a common zero, and,
if present, determine its stability. We vary $\alpha$ and $b$ in the 
intervals $-1\le \alpha \le 5$ and $2\le b \le 20$: in practice, we consider
133 different resummations.
In the $3d$-$\overline{\rm MS}$ scheme a FP is found in 91/133 cases;
it is stable in 54/133 cases. In the MZM scheme only 67 resummations 
find a FP; only in 61 cases is the FP stable. The evidence for a stable 
FP is not overwhelming: in both cases a stable FP is observed in less than 
50\% of the resummations. It is also interesting to look for the 
scatter of the estimates of the FP, see Fig.~\ref{cubico}. The estimate 
of the FP position varies significantly ($u^*$ changes by a factor of 2 in the 
$3d$-$\overline{\rm MS}$ scheme and by 20\% in the MZM scheme)
and thus it is not clear how much one can trust the estimates. 
Thus, contrary to the claim of Ref.~\cite{DHIMT-06}, 
the analysis of the perturbative series does not provide compelling evidence for
the existence of a new FP in cubic models with $v < 0$ and $N=3$. 
The evidence is even worse for larger values of $N$.

\begin{figure}[tb]
\centerline{\psfig{width=10truecm,angle=-90,file=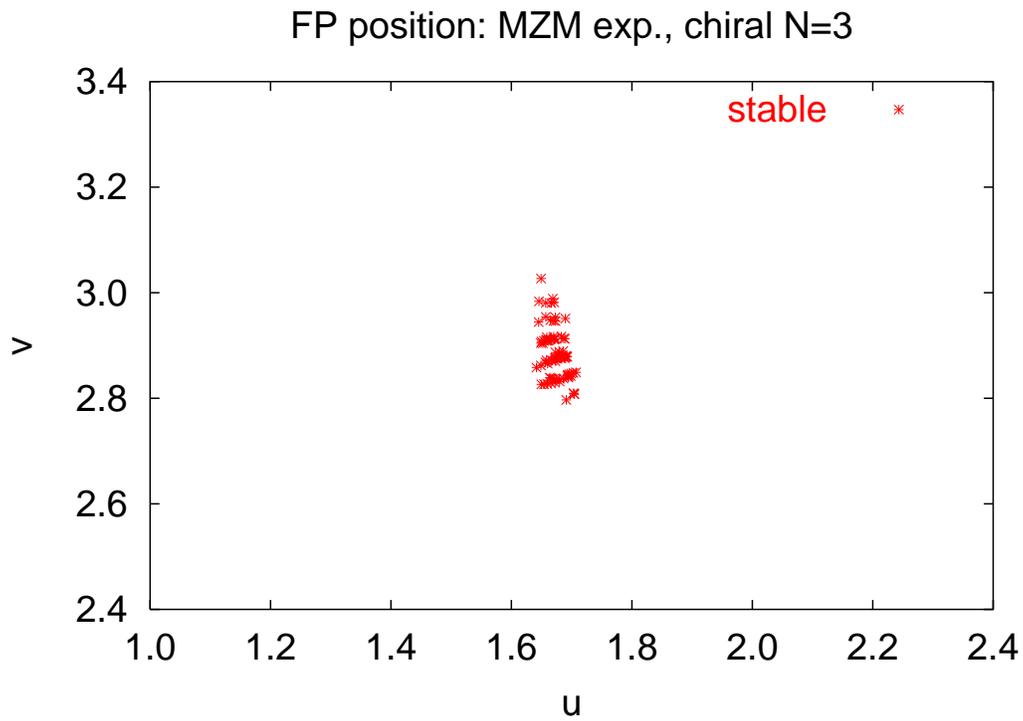}}
\vspace{2mm}
\caption{Perturbative results for the chiral model with $N=3$
and $v > 0$: Fixed-point position in the MZM scheme. Note that the 
horizontal scale is analogous to that used to show the
$3d$-${\overline{\rm MS} }$ results for the cubic model.
All FPs are stable.}
\label{chirali}
\end{figure}

In Ref.~\cite{DHIMT-06} it was claimed that the cubic-model results show 
{\em similar convergence properties} as in the frustrated case. 
This statement, given without any quantitative comparison, is completely
unjustified. For instance, in the MZM case, by using the same analysis reported
above one finds a FP in 121 cases; the FP is always stable.
While in the cubic case a stable FP is observed in less than 50\% 
of the cases, in the chiral case a stable FP is observed in 92\% of the 
cases---not a negligible difference! 
The difference between the two cases is even better understood if one compares 
the FP position, see Fig.~\ref{chirali}. Differences are so evident, that no
additional comment is needed!

\vskip 0.3truecm
{\em Reconciling the different approaches.}
To conclude this Comment we would like to go back to the original motivation 
of Ref.~\cite{DHIMT-06} (see also Ref.~\cite{DMT-04}), the discrepancy between
the perturbative results and those obtained in 
Refs.~\cite{TDM-00,TDM-03,DMT-04}, where the FRG approach was used at order 
$\partial^2$ (first order of the derivative expansion). 
In Refs.~\cite{TDM-03,DMT-04} it was noted that,
though no FP point was observed, there was a region in which 
the $\beta$ functions were small, giving rise to a 
very peculiar slowing down of the renormalization-group flow.
Correspondingly, the integration of the flow equations
led generically to good power laws ({\em pseudoscaling}) 
for the interesting thermodynamical observables. 
The presence of quasi-scaling in the FRG calculations 
suggests a way of reconciling the two approaches: 
it may be possible that by improving the approximations in 
the FRG calculation, this approximate 
scaling may turn into a real one with the presence of a true fixed point.
In other words the differences may simply be due to the crudeness of the 
approximations used in Refs.~\cite{TDM-00,TDM-03,DMT-04}. 
This interpretation is supported by the fact that 
the effective exponents obtained by using the FRG
are very close to those obtained by using 
perturbative field theory. For instance, for $N=3$ Ref.~\cite{DMT-04}
predicts (see Table XI) 
\[
\nu \approx 0.54, \qquad\qquad \beta \approx 0.29
\]
to be compared with the MZM results 
\[
\nu = 0.55(3), \qquad\qquad \beta = 0.30(2).
\]
It is worth mentioning that 
a similar phenomenon is 
observed in the two-dimensional XY model \cite{Wetterich}. 
This model shows a line of FPs
in the low-temperature phase below the Kosterlitz-Thouless transition.
FRG calculations at order $\partial^2$ do not observe this 
FP line: the $\beta$-function never vanishes in the 
low-temperature phase. Nonetheless, a clear signature of the 
presence of the line of FPs is present: the $\beta$ functions are very small 
and quasi-scaling is observed with good accuracy in the 
whole low-temperature phase. The same phenomenon
may occur in chiral models.

\vskip 0.3truecm
{\em Conclusions.} The perturbative field-theory results of 
Refs.~\cite{PRV-01,CPPV-04} are consistent with theory and with all
available experimental and Monte Carlo calculations, providing 
a consistent scenario for the critical behavior of chiral systems. 
They only disagree with FRG calculations at order $\partial^2$. 
Note, however, that, even if they do not 
find a stable FP, nonetheless they observe quasiscaling both 
for $N=2$ and $N=3$. 
One may conjecture that this quasiscaling turns into real scaling 
(with the existence of a stable FP) by improving the approximations.
This would reconcile the two approaches.

\end{document}